\documentclass[showpacs,preprintnumbers,amsmath,amssymb]{revtex4}

\allowdisplaybreaks
\usepackage{graphicx}

\input epsf
\usepackage{dcolumn}

\begin{document}
\allowdisplaybreaks
\title{One method for constructing exact solutions of equations of two-dimensional hydrodynamics of an incompressible fluid
}
\author{A.V. Yurov (*), A.A. Yurova (*,**)}%
\email{artyom_yurov@mail.ru}

{%
\affiliation{%
*The Theoretical Physics
Department, Baltic Federal University of I. Kant,
 Al.Nevsky St. 14, Kaliningrad 236041, Russia\\
** The  Mathematics Department,
Kaliningrad State Technical University,
236000, The Soviet prospectus, 1, Kaliningrad, Russia
\\
 }
\date{\today}
\begin{abstract}
We propose a simple algebraic method for constructing exact solutions
of equations of two-dimensional hydrodynamics of an incompressible fluid.
The problem reduces to consecutively solving three linear partial differential
equations for a nonviscous fluid and to solving three linear partial differential
equations and one first-order ordinary differential equation for a viscous fluid.
\end{abstract}


\maketitle

\section{\label{sec:level1}Introduction}
\allowdisplaybreaks

 The dynamics of incompressible viscous fluid flows is described by
the Navier-Stokes (NS) equation. In the regime of large Reynolds numbers
(Re), turbulence arises, presenting one of the most important unsolved
problems in theoretical and mathematical physics. Formally, letting Re tend to infinity
(or the viscosity tend to zero), we obtain the Euler equations. Whose
mathematical analysis is still more difficult than the investigation of the
NS equation. And the NS equations themselves can be regarded at large
values of ${\rm Re}$ as singularly perturbed Euler equations.The results by Kato
show that the two-dimensional NS equations are globally defined in
$C^{0}\left([0,\infty];\,H^s(R^2)\right)$, and that for $s>2$  and
$0<T<\infty$ the "weak solution" of the two-dimensional NS equations
tends to the solution of the two-dimensional Euler equation in
$C^{0}\left([0,T];\,H^s(R^2)\right)$
\cite{Kato}. In turn, the three-dimensional NS equation is locally defined in
$C^{0}\left([0,\tau];\,H^s(R^2)\right)$, and for $s>5/2$ its " weak
solutions" are approximated by solutions of the three-dimensional Euler
equations in $C^{0}\left([0,\tau];\,H^s(R^3)\right)$ when  $\tau$
is determined by the initial data (by the norm) and by the external forces
\cite{Kato1}, \cite{Kato2}. The nonviscous limit was intensively
studied in \cite{Constantin}, \cite{Wu}. The mathematical investigation
of the NS and Euler equations is undoubtedly one of the most important problems
of mathematical physics. The are now enormously many publications devoted to various
aspects of these equations. Among the most important results, we can mention
thee proof of the Hamiltonian character  of the two-dimensional
Euler equations by Arnold
 \cite{Arnold} and the investigation results on the symplectic structure
of these equations \cite{Marsden}. In
\cite{Friedlander1} and \cite{Friedlander2} the Lax representation
was found for the two-dimensional Euler equation written in the Euler
variables. The Lax representation was later also constructed in the
Euclidean variables \cite{2525}.

In this paper, we describe an astonishingly simple, but effective,
method for constructing exact solutions of equations of two-dimensional
hydrodynamics of an incompressible fluid. This method is applicable to
both nonviscous and viscous fluids (accordingly, to the Euler and NS equations).
The paper has the following structure. In Sec.2, we demonstrate this method for
the two-dimensional Euler equations describing nonviscous incompressible fluid
flows. We give some examples of exact solutions in Sec.3. In Sec.4, we generalize
the above mentioned method to the case of viscous two-dimensional flows. We
summarize the presented results in Sec.5.

\section{Two-dimensional Euler equations}
We consider the flow of a nonviscous incompressible two-dimensional
fluid. The two velocity components
$v_x$ and $v_y$ are expressed in terms of the stream function $\psi=\psi(t,x,y)$
\begin{equation}
v_x=\frac{\partial\psi}{\partial y},\qquad
v_y=-\frac{\partial\psi}{\partial x}, \label{Eul-1}
\end{equation}
as a result of which the continuity equation ${\rm div}{\vec
v}=0$. In these variables, the two-dimensional Euler equation
assumes the form \cite{Landau}:
\begin{equation}
\frac{\partial\triangle\psi}{\partial
t}+\frac{\partial\psi}{\partial
y}\frac{\partial\triangle\psi}{\partial
x}-\frac{\partial\psi}{\partial
x}\frac{\partial\triangle\psi}{\partial y}=0, \label{Eul-Euler}
\end{equation}
where $\triangle$ is the two-dimensional Laplacian.

Equation (\ref{Eul-Euler}) is a nonlinear  equation that is not one
of the so-called integrable equations (despite the existence of the Lax
representation). Nevertheless,it is possible to develop a procedure for
constructing exact solutions for it. More precisely,there exist transformations
that allow finding exact solutions of Eq.
(\ref{Eul-Euler}) that describe nonpotential (i.e.,eddying) fluid flows
proceeding from solutions describing potential flows. the crux of the matter
is expressed by the following theorem.
\newline
\newline
{\bf Theorem 1.} Let $\psi(x,y)$ be a harmonic function in a domain $D$,
i.e., $\triangle\psi=0$  where $\triangle$ is the two-dimensional Laplacian.
Let $F(t,x,y)$ be a solution of the overdetermined system of linear
differential equations
\begin{equation}
\begin{array}{cc}
\displaystyle{ \triangle F=kF},\\
\\
\displaystyle{\frac{\partial F}{\partial
t}=\frac{\partial\psi}{\partial x}\frac{\partial F}{\partial
y}-\frac{\partial\psi}{\partial y}\frac{\partial F}{\partial
x}-\frac{d\ln k}{dt}F}, \label{Eul-Backlund1}
\end{array}
\end{equation}
where $k(t)$ is a function of time. Then the function
\begin{equation}
\psi_1=\psi+F, \label{Eul-Backlund}
\end{equation}
satisfies in $D$ Eq. (\ref{Eul-Euler}), i.e.
$$
\frac{\partial\triangle\psi_1}{\partial
t}+\frac{\partial\psi_1}{\partial
y}\frac{\partial\triangle\psi_1}{\partial
x}-\frac{\partial\psi_1}{\partial
x}\frac{\partial\triangle\psi_1}{\partial y}=0.
$$
The theorem is proved by direct calculation. In what follows, formulas
(\ref{Eul-Backlund1}) and
(\ref{Eul-Backlund}) are called the ''dressing'' procedure for the  $D=2$
Euler equations.
\newline
{\bf Remark 1.} The harmonic function $\psi$ is regarded as being
dependent on two variables,i.e. $\psi=\psi(x,y)$. But it can also be assumed
to depend on $t$ as a parameter. It can be easily verified that the
theorem also holds in this case.
\newline
{\bf Remark 2.} Transformation (\ref{Eul-Backlund}) resembles the Darboux
transformation (DT) used in the theory of integrable systems \cite{BLP}. Indeed,the essence
of the DT consists in determining a solution $\Psi$ for the Lax pair with a
given ''incipient'' potential (which in turn is a solution of the nonlinear equation
under investigation) and subsequently using $\Psi$ to find new potentials.

The similarity between the DT and the transformation described above is
obvious. Indeed, as an intermediate step, we must solve two linear equations
(\ref{Eul-Backlund1}) with variable coefficient depending on the harmonic
function $\psi$, which can be regarded as an ''incipient'' potential
because it satisfies Eq.(\ref{Eul-Euler}) and describes a plane potential
flow (this is a stationary flow if $\psi$ is independent of $t$). In this
case, the new stream function $\psi_1$ in (\ref{Eul-Backlund}) describes
a nonstationary eddying flow of fluid.
Nevertheless, this is not a Darboux transformation, because nonlinear system
(\ref{Eul-Backlund1}) is not a  $[L,A]$ pair for Euler equation
(\ref{Eul-Euler}). The compatibility condition for Eqs.
(\ref{Eul-Backlund1}) has the form
\begin{equation}
\triangle\theta-k\theta=\frac{d k}{dt} F, \label{sovm}
\end{equation}
where
$$
\theta=\frac{\partial\psi}{\partial x}\frac{\partial F}{\partial
y}-\frac{\partial\psi}{\partial y}\frac{\partial F}{\partial x}.
$$
Using Eqs. (\ref{Eul-Backlund1}) and the fact that $\psi$ is a
harmonic function,we can rewrite Eq. (\ref{sovm}) in a different form,
\begin{equation}
\left(\frac{\partial^2\psi}{\partial
x^2}-\frac{\partial^2\psi}{\partial y^2}\right)\frac{\partial^2
F}{\partial x\partial y}+\left(\frac{\partial^2F}{\partial
y^2}-\frac{\partial^2F}{\partial x^2}\right)\frac{\partial^2
\psi}{\partial x\partial y}=\frac{1}{2}\frac{d k}{dt} F.
\label{sovm1}
\end{equation}

But system (\ref{Eul-Backlund1}) should not regarded as the Lax
pair for Eqs. (\ref{sovm}) or (\ref{sovm1}). The point is that a
pair of linear equations can be regarded as a Lax pair only if their
compatibility condition has the form of a nonlinear equation for the
potentials.In other words, the ''wave function''(whose role in the
case under consideration would be played by the function $F$), which is
merely an auxiliary expression, must not enter the compatibility condition.
The situation we deal with is quite different because $F$ is contained
in Eq.(\ref{sovm1}). Of course, we can take a next step in finding the
compatibility condition for Eqs. (\ref{sovm1}) and (\ref{Eul-Backlund1}),
although no elimination of the function $F$ is possible even then.We do not
discuss the interesting equation of whether this procedure can be stopped
after only finitely many equations are written.
\newline
{\bf Remark 3.} System (\ref{Eul-Backlund1}) involves an auxiliary
time function $k=k(t)$. Its expression is not fixed but is
determined by the form of the harmonic function $\psi$. In particular,
is not compatible for all harmonic functions $\psi$ system
(\ref{Eul-Backlund1})  i.e., for an arbitrary function  $\psi$,
there can be no function $k(t)$ such that system
(\ref{Eul-Backlund1}) has a solution.

We show that this elementary approach unexpectedly turns out to
be rather effective for constructing exact solutions of the
two-dimensional Euler equation.
\section{Examples of exact solutions of the Euler equation}

Wi consider a harmonic function of the form
$\psi_{_N}=X_{_N}A_{_N}Y_{_N}$ where
\begin{equation}
X_{_N}=\left(\begin{array}{ccccc} 1&x&x^2&...&x^N
 \end{array}\right),\qquad Y_{_N}=\left(\begin{array}{cccc}
1\\
y\\
y^2\\
\vdots\\
y^N \end{array}\right), \label{XY}
\end{equation}
$A_{_N}$ is a constant $(N+1)\times(N+1)$ matrix with elements
$a_{ik}$, and $i$ and $k$ range from zero to $N$. The expression
for the matrix $A_{_N}$ is found by substituting $\psi_{_N}$ in
the equation $\triangle\psi_{_N}=0$. We present several examples for
different values of $N$:
$$
A_1=\left(\begin{array}{cc} a_{00}&a_{01}\\
a_{10}&a_{11}
\end{array}\right),\qquad
A_2=\left(\begin{array}{ccc} a_{00}&a_{01}&a_{02}\\
a_{10}&a_{11}&0\\
-a_{02}&0&0
\end{array}\right),
$$
$$
A_3=\left(\begin{array}{cccc} a_{00}&a_{01}&a_{02}&a_{03}\\
a_{10}&a_{11}&-3a_{30}&a_{13}\\
-a_{02}&-3a_{03}&0&0\\
a_{30}&-a_{13}&0&0
\end{array}\right),\qquad
A_4=\left(\begin{array}{ccccc}
a_{00}&a_{01}&a_{02}&a_{03}&a_{04}\\
a_{10}&a_{11}&-3a_{30}&a_{13}&0\\
-a_{02}&-3a_{03}&-6a_{04}&0&0\\
a_{30}&-a_{13}&0&0&0\\
a_{04}&0&0&0&0
\end{array}\right),
$$
$$
A_5=\left(\begin{array}{cccccc}
a_{00}&a_{01}&a_{02}&a_{03}&a_{04}&a_{05}\\
a_{10}&a_{11}&-3a_{30}&a_{13}&5a_{50}&-\frac{3 a_{33}}{10}\\
-a_{02}&-3a_{03}&-6a_{04}&-10a_{05}&0&0\\
a_{30}&-a_{13}&-10a_{50}&a_{33}&0&0\\
a_{04}&5a_{05}&0&0&0&0\\
a_{50}&-\frac{3 a_{33}}{10}&0&0&0&0
\end{array}\right),
$$
etc. The number $M(N)$  of free parameters(i.e.,the number of
independent elements in the matrices $A_{_N}$) is determined as follows:
if $N$  is odd,then $M(N)=2(N+1)$, and if  $N$  is even,then $M(N)=2N+1$.
Because the stream function serves only as an auxiliary means for calculating
the velocity field and vanishes under differentiation with respect to the
spatial variables $a_{00}$,we can set $a_{00}=0$ without loss of generality,
which is precisely implied in what follows. The other entries should be
regarded as functions of time, i.e., $a_{ij}=a_{ij}(t)$. Substituting $\psi_{_N}$
in system (\ref{Eul-Backlund1}) and solving it, we extract the function $F$,
after which the corresponding function $\psi_{1,{_N}}$ is
calculated using formula (\ref{Eul-Backlund}).

We demonstrate this approach for some specific examples. Let $N=1$. In
this case, $a_{11}=0$ should be chosen.It is convenient to parameterize
$a{01}(t)$ and $a_{10}(t)$ as
$$
a_{01}(t)=A(t)\sin\alpha(t),\qquad a_{10}(t)=A(t)\cos\alpha(t),
$$
where $\alpha(t)$ and $A(t)$ are arbitrary real functions. After a simple
calculation, we finally obtain
\begin{equation}
\psi_1=A(t)\left(y\sin\alpha(t)+x\cos\alpha(t)\right)+\sum_{i=1}^n\left(c_i\sin
g_i+c'_i\cos g_i\right), \label{resh-1}
\end{equation}
where $g_i=x\cos\phi_i+y\sin\phi_i+f_i(t)$, $\phi_i$,
$c_i$, and $c'_i$ are arbitrary constants, and $f_i(t)$ is given
by the integral formula
$$
f_i(t)=f_i(0)+\int dt A(t)\sin\left(\phi_i-\alpha(t)\right).
$$
In this case, $k$ is simply a negative constant~\footnote
{This is a dimensional constant, namely, $[k]=1/$m${}^2$. To
avoid ambiguity, we note that the arguments of the sine and
cosine functions in (\ref{resh-1}) are multiplied by $\sqrt{-k}$.}, which
is set to minus unity without loss of generality. Expression (\ref{resh-1})
can be straightforwardly substituted in (\ref{Eul-Euler}) to
to ensure that we have obtained a solution of the Euler equation
parameterized by two arbitrary functions $A(t)$ and
$\alpha(t)$. we take $n=1$, $A(t)=\mu/\cosh^2\nu t$, $\alpha={\rm
const}$, $c'_1=0$, $c_1=c$, $\phi_1=\phi$, and $f_1(0)=0$. In this
special case, the solution takes the form
\begin{equation}
\psi_1=\frac{\mu}{\cosh^2\nu
t}\left(y\sin\alpha+x\cos\alpha\right)+c\sin\left[\kappa\left(x\cos\phi+y\sin\phi+\frac{\mu}{\nu}\sin(\phi-\alpha)\tanh\nu
t\right)\right], \label{chast}
\end{equation}
where $\kappa=\sqrt{-k}$, $k<0$ is introduced. Figure 1-3
shows the velocity fields for the corresponding flow at the
instants $t=0$, $t=1$, and $t=50$.Here, $\alpha=0$, $\phi=\pi/4$, and
the other (dimensional) parameters are set to unity.
\begin{figure}[ht]
\begin{center}
\epsfxsize=10cm\epsffile{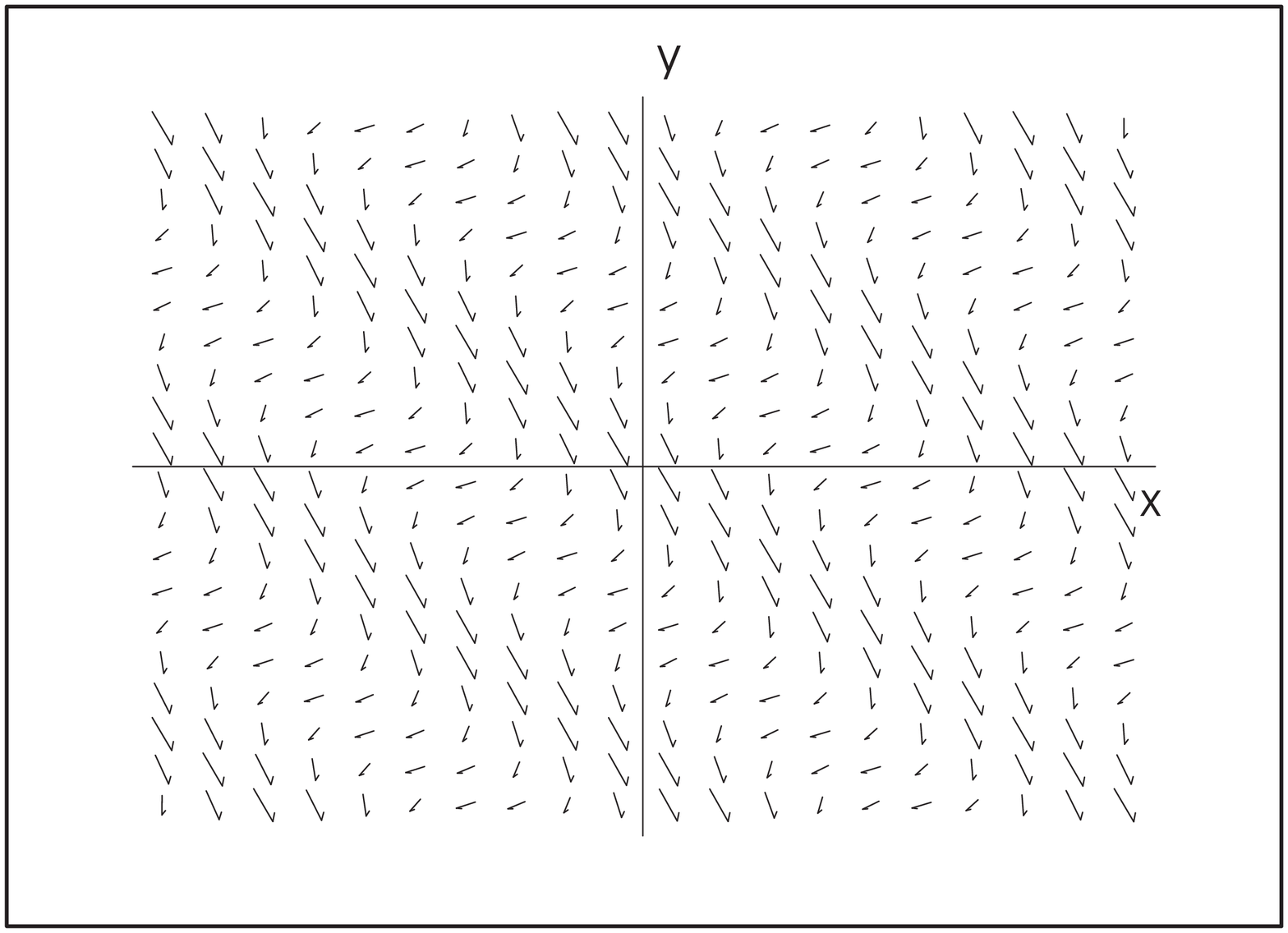} \caption{Graph of the velocity field
for $N=1$, at the instants $t=0$ in the domain $-10<x,y<10$. }
 \label{singplot}
\end{center}
\end{figure}
\begin{figure}[ht]
\begin{center}
\epsfxsize=10cm\epsffile{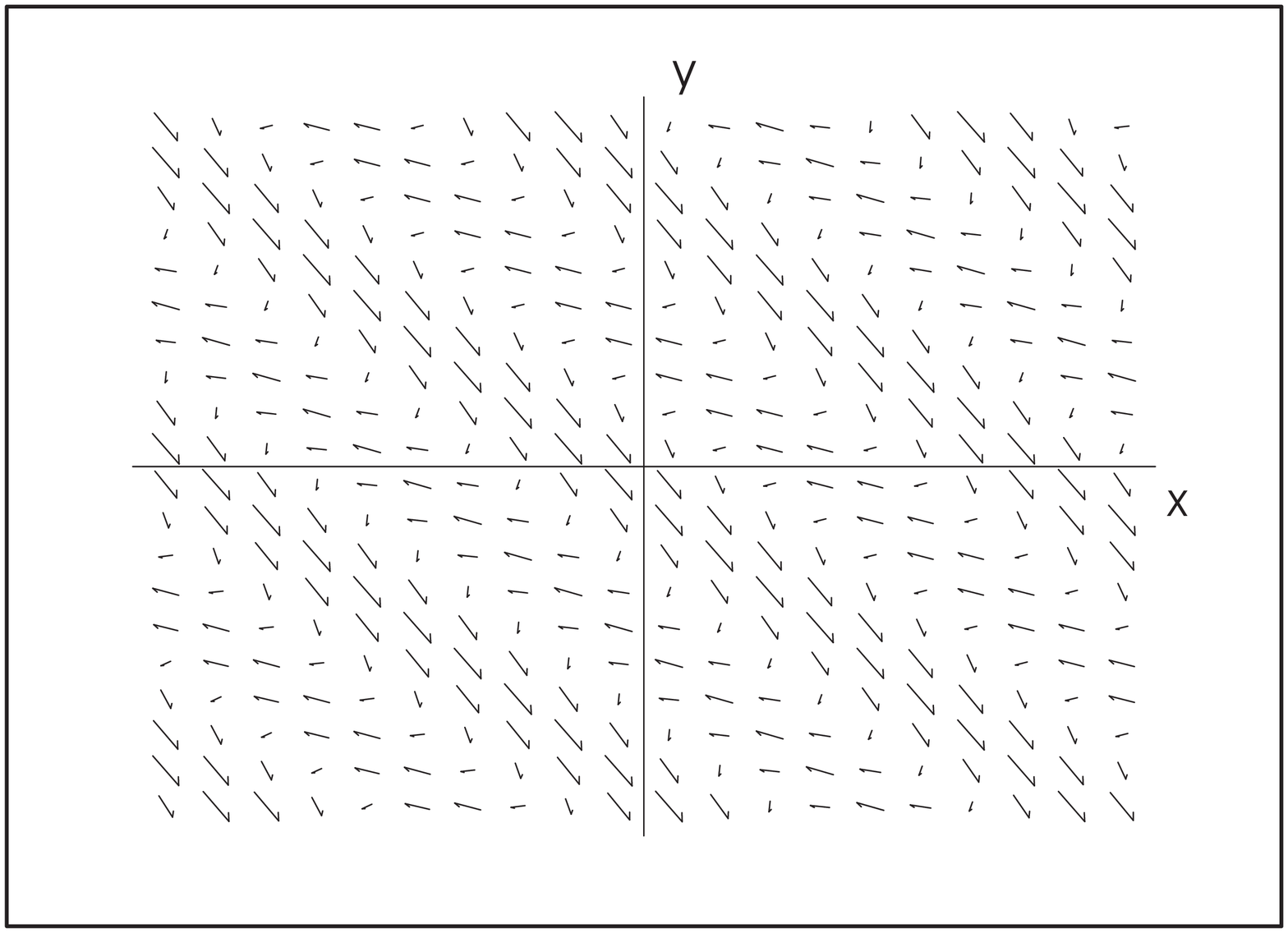} \caption{Graph of the velocity field
for $N=1$, at the instants $t=1$ in the domain $-10<x,y<10$. }
 \label{singplot}
\end{center}
\end{figure}
\begin{figure}[ht]
\begin{center}
\epsfxsize=10cm\epsffile{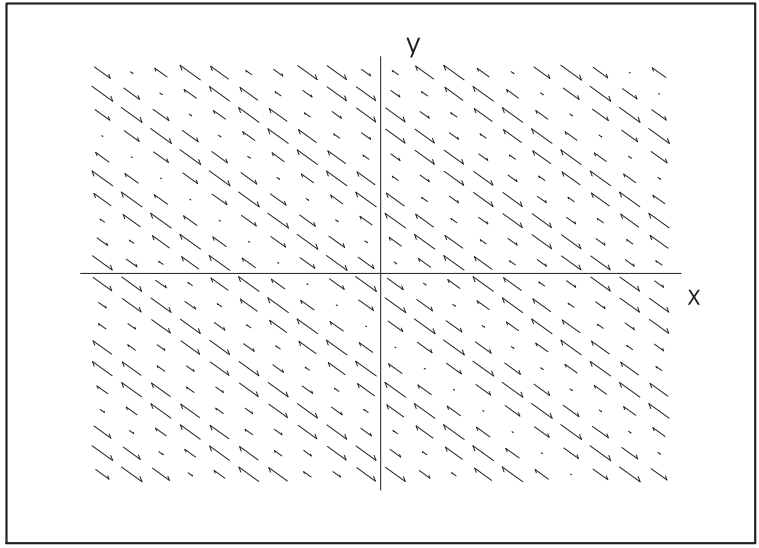} \caption{Graph of the velocity field
for $N=1$, at the instants $t=50$ in the domain $-10<x,y<10$. }
 \label{singplot}
\end{center}
\end{figure}
The resulting solution qualitatively resembles the so-called ''exulton'',
a specific solution of the nonlinear Schrodinger equation \cite{Exul}.
An exulton is a rational perturbution that appears on the background
of a pereodic wave, increases, and then rapidly disappears~\footnote
{From the Latin exultare (exsultary), to leap up frequently or rejoice.}.
Solution (\ref{chast}) behaves similarly, namely, at negative times sufficiently
large in absolute value, the flow looks qualitatively the same as Fig.3.
As $t$ aspiring to zero at the left, the first term in solution
(\ref{chast})begins to dominate, which distorts this pattern. The maximum
distortion occurs at $t=0$ (see Fig. 1).After that, the first term disappears
(\ref{chast}) exponentially fast, and we again have original (although somewhat
displaced) pattern shown in Fig. 3.

We now consider the case $N=2$. Assuming that $a_{01}$, $a_{10}$,
$a_{11}$ and $a_{02}$ are function of time, we seek $F$ in the form
$F=\xi(t)\sin(a(t) x+b(t) y+c(t))$. Solving Eqs.
(\ref{Eul-Backlund1}) and using function  (\ref{Eul-Backlund}) we obtain
\begin{equation}
\psi_1=a_{02}(t)\left(y^2-x^2\right)+a_{11}(t)xy+a_{01}(t)y+a_{10}(t)x+\frac{\xi_0\sin(a(t)
x+b(t) y+c(t))}{a^2(t)+b^2(t)}, \label{resh-2}
\end{equation}
where $\xi_0$=const, and the functions $a(t)$, $b(t)$ and $c(t)$ are found from
system of ordinary differential equations
\begin{equation}
\begin{array}{l}
\displaystyle{\frac{da(t)}{dt}=-a_{11}(t)a(t)-2a_{02}(t)b(t),\qquad
\frac{db(t)}{dt}=a_{11}(t)b(t)-2a_{02}(t)a(t)},\\
\\
\displaystyle{\frac{dc(t)}{dt}=-a_{01}(t)a(t)+a_{10}(t)b(t)}.
\end{array}
\label{difur}
\end{equation}
The functions $a_{ij}(t)$ can now be chosen arbitrarily. At the
next step,we substitute them in system (\ref{difur}) and solve it for
$a(t)$, $b(t)$, and $c(t)$. Finally, substituting the resulting
functions in (\ref{resh-2}) we obtain the desired solution.

As a simple example, we set $a_{11}=$const and
$a_{02}$=const. Calculating as described above, we find
$$
a(t)=a_0\cosh\omega t,\qquad
b(t)=-\frac{a_0}{2a_{02}}\left(a_{11}\cosh\omega
t+\omega\sinh\omega t\right),
$$
where $a_0$ is the integration constant and
$\omega=\sqrt{a_{11}^2+4a_{02}^2}$. For $c(t)$, it is convenient to
set $a_{10}(t)b(t)=a_{01}(t)a(t)$, which gives $c$=const and we can
choose $c=0$ without loss of generality. We note that solution
(\ref{resh-2}) is not singular, and we have  $k(t)=-a^2(t)-b^2(t)$.

The behavior of the velocity field at $t=0$  and $t=50$ is shown
in Fig. 4, 5, where we have $a_{10}=a_{01}=0$, and all the other parameters
are set to zero.
\begin{figure}[ht]
\begin{center}
\epsfxsize=10cm\epsffile{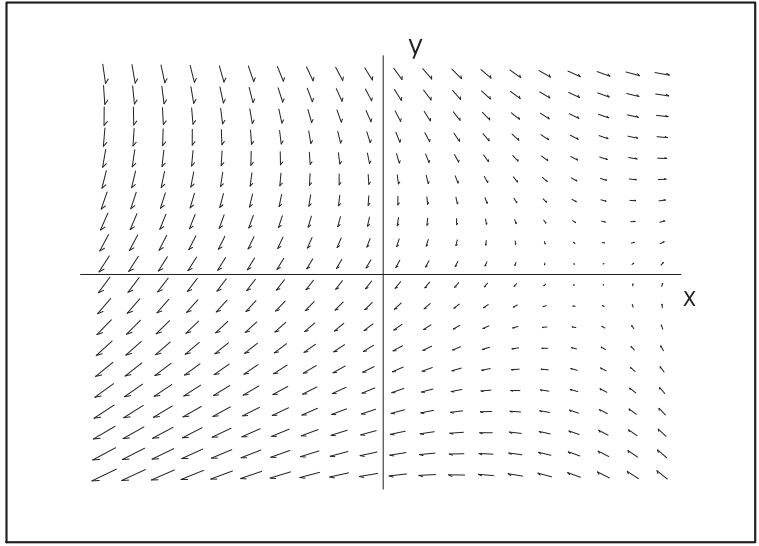} \caption{Graph of the velocity field
for $N=2$, at the instants $t=0$ in the domain $-0.5<x,y<0.5$. }
 \label{singplot}
\end{center}
\end{figure}
\begin{figure}[ht]
\begin{center}
\epsfxsize=10cm\epsffile{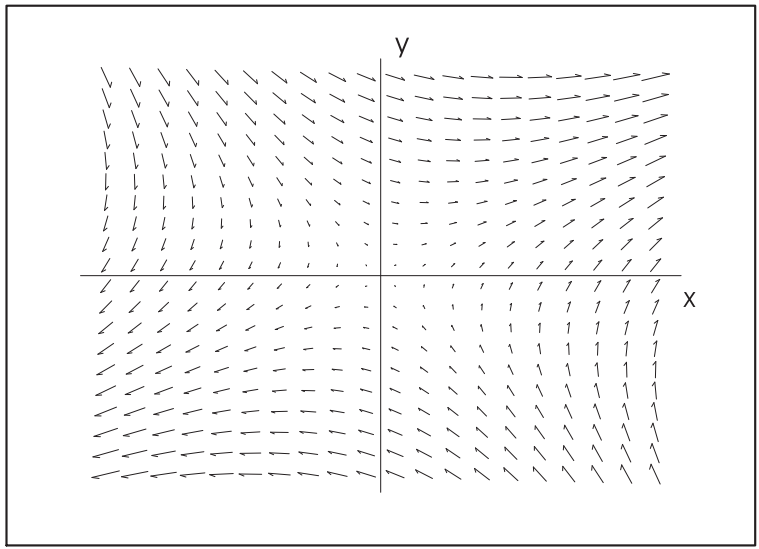} \caption{Graph of the velocity field
for $N=2$, at the instants $t=50$ in the domain $-0.5<x,y<0.5$. }
 \label{singplot}
\end{center}
\end{figure}
We note  that in contrast to the  previos solution, at times
$t$  with a large absolute value, the first term dominates instead
of the second. Furthermore, for large values of $x$ and $y$, the
first term in expression (\ref{resh-2}) also dominates.
Because the distortion of the structure occurs only in the neighborhood
of the origin as $t\to 0$, the small spatial domain
$-0.5<x,y<0.5$. is taken in Fig.2; the indicated changes would
not be noticeable at the scales chosen in Fig.1.

Finally, we note that the same method can also be used to construct solution
describing potential flows. For example, taking $N=3$ and $k=0$, we
can obtain the solution  of the Euler equation
$$
\begin{array}{cc}
\displaystyle{
\psi_1=a_{01}y+a_{02}y^2+a_{03}y^3+a_{10}x+a_{11}xy-3a_{30}xy^2+a_{13}xy^3-a_{02}x^2-3a_{03}x^2y+}\\
\\
\displaystyle{+a_{30}x^3-a_{13}x^3y+{\rm
e}^{ax+by+c}\left(A\sin(bx-ay-{\tilde c})+B\cos(bx-ay-{\tilde
c})\right),}
\end{array}
$$
where all coefficients except $A$ and $B$ are arbitrary functions
of time.It can be easily shown that $\triangle\psi_1=0$.
Of course, two-dimensional potential flows can be described using
the powerful method based on introducing a complex potential
\cite{Landau}, and using the technique described above therefore seems
inconvenient in this case. The advantages of our approach for constructing
exact solutions in the case of two-dimensional Euler equations
become obvious only when applied to nonstationary eddying flows of fluid.

\section{Two-dimensional flow of an incompressible viscous fluid}
However astonishing, the technique described above can be easily generalized
to the case of a two-dimensional flow of an incompressible viscous fluid described
by the equations
\begin{equation}
\frac{\partial\triangle\psi}{\partial
t}+\frac{\partial\psi}{\partial
y}\frac{\partial\triangle\psi}{\partial
x}-\frac{\partial\psi}{\partial
x}\frac{\partial\triangle\psi}{\partial y}=\nu\triangle^2\psi,
\label{Navie}
\end{equation}
where $\nu$ is the kinematic viscosity. Namely, the following theorem holds.
\newline
\newline
{\bf Theorem 2.} Let $\psi$ be a harmonic function in a domain $D$, i.e.,
$\triangle\psi=0$. Let $F(t,x,y)$ be a solution of the
overdetermined system of linear differential equations
\begin{equation}
\begin{array}{cc}
\displaystyle{ \triangle F=u(t)F},\\
\\
\displaystyle{\frac{\partial F}{\partial
t}=\frac{\partial\psi}{\partial x}\frac{\partial F}{\partial
y}-\frac{\partial\psi}{\partial y}\frac{\partial F}{\partial
x}+U(u)F}, \label{Navie-Backlund1}
\end{array}
\end{equation}
where $u(t)$ satisfies the ordinary differential equation
\begin{equation}
\frac{du(t)}{dt}+uU(u)=\nu u^2, \label{Navie-ut}
\end{equation}
and $U(u)$ is an arbitrary function of the argument. Then the
function $\psi_1=\psi+F$ in $D$ satisfies Eq. (\ref{Navie}):
$$
\frac{\partial\triangle\psi_1}{\partial
t}+\frac{\partial\psi_1}{\partial
y}\frac{\partial\triangle\psi_1}{\partial
x}-\frac{\partial\psi_1}{\partial
x}\frac{\partial\triangle\psi_1}{\partial y}=\nu\triangle^2\psi_1.
$$
As in the case of Theorem 2, Theorem 1 is proved by a straightforward
calculation.

As the simplest example, we consider the ''dressing'' on the zero background,
i.e.,$\psi=0$. we choose $U(u)=\nu u$. As follows from Eq.
(\ref{Navie-ut}), expression $u$ is a constant. Solving system
(\ref{Navie-Backlund1}), we obtain
$$
\displaystyle{\psi_1=K {\rm
e}^{-\nu\left(a^2+b^2\right)t}\left(C_1\sin(ax+by)+C_2\cos(ax+by)\right)},
$$
where $K$, $a$, $b$, and $C_{1,2}$ are arbitrary constants.
Of course,this simple and well-known solution, which we present here only
to demonstrate the workability of the method.

We now consider case $N=1$. By analogy with expression (\ref{resh-1}), we can
construct a solution of Eq. (\ref{Navie}) in the form
\begin{equation}
\psi_1=A(t)\left(y\sin\alpha(t)+x\cos\alpha(t)\right)+\xi_1(t)\sin
g+\xi_2(t)\cos g, \label{Navie-resh2}
\end{equation}
where
$$
g=R\left(x\cos\phi+y\sin\phi+f(t)\right),\qquad U(u)=\nu u,\qquad
u=-R^2={\rm const},
$$
$A(t)$, $\alpha(t)$, and $f(t)$ are arbitrary function, $\phi$
is an arbitrary constant, and the functions $\xi_1(t)$, and $\xi_2(t)$
are solutions of the system of ordinary differential equations
$$
\begin{array}{c}
\displaystyle{ \frac{d\xi_1(t)}{dt}=-\nu R^2\xi_1+R\left(\frac{d
f(t)}{dt}-A(t)\sin(\phi-\alpha(t))\right)\xi_2},\\
\\
\displaystyle{ \frac{d\xi_2(t)}{dt}=-\nu R^2\xi_2-R\left(\frac{d
f(t)}{dt}-A(t)\sin(\phi-\alpha(t))\right)\xi_1}.
\end{array}
$$
Of course, by analogy with expression (\ref{resh-1}), a superposition of
arbitrarily many sine and cosine functions can be constructed instead of
(\ref{Navie-resh2}). This observation is a consequence of the linearity of
Eqs. (\ref{Navie-Backlund1}) and of the linearity of the transformed
function $\psi$ in the spatial variables.
In special case $df(t)/dt=A(t)\sin(\phi-\alpha(t))$, we obtain
$$
\psi_1=A(t)\left(y\sin\alpha(t)+x\cos\alpha(t)\right)+{\rm
e}^{-\nu R^2t}(\xi_1(0)\sin g+\xi_2(0)\cos g),
$$
where $\xi_{1,2}(0)$ is the integration constant. Comparing this
(\ref{resh-1}) and (\ref{chast}), we see that including viscosity, as
should be expected, leads to an additional exponential factor descibing
the dissipation.

\section{Conclusion}

The equations of two-dimensional hydrodynamics of an incompressible
fluid thus admit a simple algebraic method for constructing exact
solution. Here, we confined ourself to demonstrating the simplest
solution for the cases of nonviscous and viscous fluid separately.
There is no doubt that the suggested technique can be used to construct
many much more complicated and physically interesting solutions.

The most interesting open equation is using the ''dressing
method'' to solve boundary problems. For example, we consider
Eqs. (\ref{Navie-Backlund1}) and (\ref{Navie-ut}). If we solve  the
two equations $\triangle\psi=0$ and $\triangle F=u(t)F$ with given
identical boundary conditions $S$ (it is reasonable to consider the
second boundary problems,i.e.,the Neumann problem), then the solution $\psi_1$
obtained by the above mentioned method will satisfy the boundary
conditions $2S$. On the other hand, the function $F$ must  satisfy
an additional dynamical equation, which may turn out to be incompatible
with solution of the given boundary problem. Consequently, the equation
should be stated as follows. What is the class of Neumann boundary
conditions that is compatible with system (\ref{Navie-Backlund1}),
(\ref{Navie-ut}) and the equation $\triangle\psi=0$? We hope to return
to this problem in our further publications.
\newline
\newline
{\bf Acknowledgments}
\newline
\newline
The authors express their gratitude to the referee, whose valuable
remarks and suggestions undoubtedly allowed them to improve the paper.
The authors are grateful to A.K.Pogrebkov and Y.Li. One of the authors
(A.V.Yu) cordially thanks the Department of Mathematics, University
of Missouri-Columbia(USA), for the Miller's Scholar position.


\end{document}